# Navigating the Quantum Divide(s)

Ayda Gercek, *Middle East Technical University-STPS, Ankara, Turkey*
Zeki C. Seskir, *Karlsruhe Institute of Technology-ITAS, Karlsruhe, Germany*

This article explores the potential divides from emerging quantum technologies (QT) on society, covering impacts on science, technology, geopolitics, and societal structures. We aim to challenge the concept of a single "quantum divide" by offering a comprehensive view. We propose four divides: in science, technologies, between countries, and within societies. The "Quantum Divide in Science" highlights gaps among scientists and access inequalities in research communities. The second divide, "Quantum Divide in Technologies through Path-dependency," looks at technology adoption processes. The "Quantum Divide between Countries" examines adoption reasons and outcomes across countries with varying development levels. Lastly, the "Quantum Divide within Societies" addresses societal implications, one of the most discussed in literature. We propose directions to navigate these divides, noting their interconnectedness and impact on innovation ecosystems. This study guides those interested in the quantum divide, navigation strategies, and QT's influence on scientific, technological, international, and societal institutions.

**Keywords:** Geopolitics, Path-dependency, Societal impact, Quantum divide, Quantum technologies

## I. INTRODUCTION

The first quantum revolution occurred in the early 20th century and has had a profound impact on our daily lives, resulting in the development of a wide selection of impactful outcomes ranging from the computer to the nuclear weaponry. The second quantum revolution, which started in the 1980s, is expected to have similar effects before mid-century [1]. As being the cause of these revolutions, quantum mechanics brought new discoveries in the understanding of fundamental principles of nature. The first quantum revolution emerged under the leadership of Erwin Schrödinger and Werner Heisenberg in the 1920s. This revolution came to life with quantum mechanics, which introduced the wave function, Heisenberg's uncertainty principle, and the intrinsically probabilistic nature of quantum events [2]. The ideas of energy quantization, wave-particle duality, and the uncertainty principle have led to advances in various fields of science and the development of many modern technologies such as lasers, MRI scanners, and transistors [3]. With the advent of the second quantum revolution, this trend is only expected to accelerate. Simply put, these developments are creating new opportunities for exploration and expanding our understanding of the physical world [1], and our ability to shape it.

In his famous book 'The Structure of Scientific Revolutions', Kuhn suggests that in a given time there is usually one paradigm that makes up the way of experiencing, understanding, and interpreting the world. This paradigm has the best answers to the questions of the scientists at that time. During periods of 'normal science', as termed by Kuhn, scientists operate within this paradigm, engaging in problem-solving based on the established theories and practices. However, through time, the existing paradigm fails to provide adequate answers to the new questions raised, and as a result, tension in the system that calls for a revolution starts accumulating. Eventually, the existing paradigm shifts to one that has better answers and with those answers, carries a new way of understanding and living the world. A common example of this shift is given as the transition from Newtonian to Einsteinian laws of physics, as explained by Kuhn [4]. This Kuhnian conceptualization of scientific revolutions and paradigms are reformulated in the 1980s for the field of economic development via the neo-Schumpeterian quest to understand innovation and processes (and difficulties) of technological catch-up by authors such as Dosi [5] and Perez & Soete [6]. This formulation relies heavily on the distinction between invention and innovation, and the fact that "…*the space of the technologically possible is much greater than that of the economically profitable and socially acceptable*" [7, p.186]. Thus, which technology takes the lead in certain areas compared to its alternatives and for what reasons heavily relies on factors outside of the bounds of that technology's technical capabilities. Societal and economic stakeholders from government, industry, academia, and civil society all play a role in setting the conditions for adoption and trajectory of innovation. This line of thinking is aligned with the quadruple helix model of innovation [8] popularized in late-2000s, which emphasizes the importance of integrating the perspectives of all the stakeholders, including the media-based and culture-based public, compared to the previously prevalent triple helix model [9] that locates the crux of innovation in the university—industry—government relations. Hence, a technological revolution both impacts and is impacted by a wide range of stakeholders in our society. It goes beyond just the scientists working on the fundamentals or the engineers and designers who turns creates the products, all four types of stakeholders have a role to play in bringing the new paradigm.

Every technology within the constellation of QT has its revolutionary promises. Quantum computer's ability to perform some complex calculations much faster than classical computers gives it a significant computational advantage [10]. Quantum key distribution (QKD) and quantum communications offer a new level of security and novel methods in data transmission and protection against cyber threats through unbreakable encryption [11], and the groundwork for a potentially novel new infrastructure, the quantum internet [12]. There are practical applications in fields like navigation, environmental monitoring, and medical diagnosis through the use of high-precision quantum sensors [13], which can also improve precision in areas like imaging techniques, quality control, and metrology [14]. Quantum simulation offers better understanding and control over materials benefiting from quantum properties, such as superconductors, and have the potential to revolutionize energy storage and electronics through enabling more advanced materials science research [11]. Furthermore, quantum mechanics is still driving fundamental scientific discoveries, reshaping and advancing technology and research [11].

It should be noted that despite the promises made by some scientists and entrepreneurs, the tangible products of the QT with revolutionary impacts, are not anticipated in the immediate future [15]. Consequently, the element of uncertainty in the specifics of these advancements is undeniable. It is safe to argue that most research in new QT is still emerging out of their era of basic research [3], where no dominant design is established. Quantum computers of the future can use superconducting circuits, ion-traps, photonics, neutral atoms, or another modality of physical realization, all of which have different advantages and disadvantages in terms of operation speed, coherence times, and connectivity which will directly impact the application areas they will be able to bring most value to. They have different value chains, different expertise pools, and different material requirements, which results in varying configurations of necessary partnerships between companies and countries. Similarly, the future of the quantum internet is also in flux, where different quantum repeater models with or without memories, entanglement distribution methods, and even infrastructure types (fiber vs. satellite) are still in negotiation between the stakeholders, and no clear path forward is laid on the table that has a buy-in from all the stakeholders. However, regardless of the exact configuration and design of the tangible outputs in the mid-to-distant future, a constant reality is that with any scientific or technological breakthrough of this magnitude, divides can emerge over who will benefit and who will be harmed from the widespread introduction of these technologies to society.

In evolutionary economics, one way to study such situations is the path-dependency [16] approach, which is particularly useful in imagining the potential trajectories of emerging technologies, and their impacts on different stakeholders and various areas of life. Decisions of the past put a constraint on the range of decisions available and their cost today, and the decisions today will put a constraint on the range of available decisions and their cost in the future. This holds true both for a country and a company, where policy decisions and implementation for a country, and the managerial decisions and implementation for a company, cause certain paths to be prioritized over others and alter the cost structure in different periods of time. For example, a company heavily invested in ion-trap technology in the earlier stages of development in quantum computing will have a higher cost of integrating itself into the global value chain if the dominant design for quantum computers ends up being a different modality (like superconducting or photonics). There are many such decisions in the early stages of a radical innovation that has the potential of drastic impact on different stakeholders in the future, creating divides between them.

In the context of technology and society, the concept of "divide" generally refers to a significant gap that exists between different groups or regions with respect to 'access to' and 'utilization of' a particular technology or resource, which causes inequality. This term was popularized in literature to point out divides emerged in the case of the information age [17]. Therefore, the divide is mostly used to describe situations where some individuals or communities are able to access and benefit from certain technologies while others are unable to do so, leading to digital, technological, or information segregation. In our case, the concept of quantum divide corresponds to a possible gap between groups and/or countries in any conceivable area of life, mainly caused by the lack of interaction or understanding of quantum mechanics, technologies, and their possible outcomes [18], [19]. The importance and significance of bridging the gap between theoretical and practical applications of QT is already pointed out by several studies in the literature [18], [19], [20].

In previous studies, the notion of divide is mainly constructed from the perspectives of a divide between countries and within societies. They emphasize the need to address the challenges associated with QT and increase collaboration among stakeholders such as researchers, industry experts, and policymakers. In our work, we follow the line of thinking that more than one type of divide needs to be taken into account in order to talk about a comprehensive and inclusive development environment for QT [19]. In the existing literature, there is a definitional gap in the categorization and evaluation of these divides. To address this gap, we propose four different types of quantum divides arising from the four major stakeholder areas defined in the quadruple helix model [8] to be described and analyzed in this work. The benefit of evaluating the phenomenon of the "divide" under four distinct headings lies in moving away from a single-focused, one-dimensional perspective, allowing for a more comprehensive examination of events and phenomena in their interconnected contexts. This paper leans towards a scenario analysis approach. By analyzing current data and proposing trends to be followed in the future, it differentiates itself from existing studies in the literature. While prior works focus on observing QT-related divides within a single domain, this paper offers a multidimensional perspective to the analysis and projection of the "divide," contributing a unique and holistic approach that sets it apart. The quadruple helix model identifies academia, industry, governments, and civil society as the relevant stakeholder groups for a functioning innovation ecosystem, we propose a corresponding divide for each of these. These are the quantum divides in science, in technology through path-dependency [16], between countries, and within societies. Therefore, while the majority of the previous studies have taken the element of access as the main source of division and focus on divides within societies and between countries, we argue that a more nuanced approach is better suited for enabling a wider range of actions from different stakeholders within innovation ecosystems. Expanding the discussion beyond the country and society-wide scopes provides an opportunity to bring in researchers, technology developers and engineers from industry and business stakeholders, especially those with managerial responsibilities, into the debate on how to navigate these divides and how to identify the conflicting interests between different stakeholders with respect to them.

## II. WHY ARE QUANTUM TECHNOLOGIES EXPECTED TO RESULT IN A DIVIDE?

It is argued by Manuel Castells that we live in a network society [21]. Castells' theory of the Network Society, particularly concerning digital technologies, highlights the significant influence of informational networks, like the Internet, on modern society. This viewpoint suggests that networks created by major historical events—whether through innovations, inventions, or political actions—have a deep impact. For example, the advent of the internet and smartphones has drastically altered our communication methods and business operations, changing everyday life and social interactions. In such a society, the key social structures and activities are organized around electronically processed information networks. Within this context, the digital divide is argued to be not limited to the simple matter of having access to technology, the internet. It is a complex phenomenon that involves a range of factors beyond just physical availability. This line of thinking contrasts with some others. For example, Michael K. Powell, the Federal Communications Commission chairman of the time, refers to the digital divide in 2001 as calling it a "Mercedes divide". From his point of view, it is a similar case to "*wanting a Mercedes car, but not being able to have it*" [22]. In comparison, Castells argues [23] that a divide caused by technological developments has far-reaching implications for knowledge creation and destruction, as well as power dynamics. It is "a complex interaction" in a social structure that holds the sources of inequality and social exclusion (p.781). It is not as simple as wanting but not having a Mercedes car.

These divisions often arise due to the strategic actions and managerial decisions of various stakeholders with external motivations, not aiming at creating a divide. Such motivations include pursuing economic development, attaining self-sufficiency, or gaining political power in the case of countries. For example, the World Bank highlights the importance of self-sufficiency and suggests that policymakers should support the development of local data infrastructures [24]. However, it is observed and discussed in the literature [25] that for smaller countries without significant market size, such regulations can cause main providers to not invest in local infrastructure. This, in return, limits access to these services and increases costs for local businesses, preventing entrepreneurs of these countries from fully participating in the digital economy and further exacerbating the digital divide.

QT are also expected to be transformative technologies by many [15, 26, 27], for having the potential to significantly impact and disrupt various sectors, including healthcare, defense, finance, energy, and others. In these lines, the concept of quantum divide is sometimes constructed as "*challenges, between regions, between nations, and between rich and poor within nations, around creating more equitable access to quantum computing*" [19, p.3]. Furthermore, in this context accessibility is identified as a core value in the field of QT [20, p. 274], [28] and unequal access is perceived as the main cause of the quantum divide [29]. On the other hand, some conceptualizations of the quantum divide refer to it as a "knowledge gap" [18, p.242], and put the distinction between individuals engaged in QT development and those who are not [30], or even the knowledge gap between quantum scientists and society [31]. In short, the quantum divide, similar to the concept of digital divide, is conceptualized in the literature primarily associated with having access to devices and knowledge on QT or not.

It should be noted that QT are not the only set of technologies that is expected to result in a divide, divides emanating from other emerging technologies such as artificial intelligence (AI), robotics, and nanotechnology are also expected and discussed in the literature. There are recent studies [32] and reports published on the "AI Divide", such as a recent one from the United Nations and International Labour Organization, and on the "Robotics Divide" [33, 34]. Similar discussions can also be found on nanotechnology dubbed as "Nano-Divide" [35, 36] dating back to 2006. Some might argue that since we do not see a nano-divide on the level of a digital divide today, the same can be expected for QT. However, it is important to be prepared for the case if such extreme disparities arise, especially since globally more than $55 billion of mostly public investments are designated for QT [37] development. Given the revolutionary nature of QT [1], the high levels of investment globally, and a strong strategic interest from important actors from NATO [38] to the European Commission [39], it is only expected that QT might result in a divide which has far-reaching implications for knowledge creation and destruction, as well as power dynamics, impacting all

stakeholders [23].

*A. Types of Divides*

As in all emerging technologies supported by public funding, the main preferred result of those impacts can be thought to be increasing welfare in general. As has been mentioned in the previous section, there are various sectors that are expected to be affected by QT that can serve that direction. Furthermore, one of the far-reaching impacts of the QT, besides the impacts on "*(leader) countries' industrial bases and providing economic and national security benefits*," has been argued as creating jobs [29]. Beyond sectoral influences, QT can play a role in reshaping social and cultural landscapes. This prediction, while seemingly speculative, underscores the profound entanglement of technology with the fabric of society, influencing political, social, and cultural spheres far beyond a mere tool. QT embodies the values and aspirations of the communities that forge and adopt it. These effects, undoubtedly, will not manifest in the near future, but they can work their way through means such as humanities [40, 41], games [42, 43], music [44], and arts [45].

Considering that QT has been noted as having strategic importance for governments [38, 39], has disruptive potential for many of the critical sectors [29], can scientifically play a revolutionary role [1], and is already impacting the social and cultural spaces [44], we formulate the different types of divides that could potentially arise by covering the entire stakeholder base of the quadruple helix. We propose four types of divides that have the potential to emerge during the development, adoption, and usage processes of QT. These are the quantum divide in science, the quantum divide in technologies through path-dependency [16], the quantum divide between countries, and the quantum divide within societies. In the following subsections from A.I to A.IV we introduce and describe these types of divides. The discussions on how to navigate them are reserved for Section III.

*A.I. Quantum Divide in Science*

There are no specific references to the term 'scientific divide' in the existing literature. Nevertheless, we find this term suitable to characterize what we have in mind. The concept of a scientific divide can be built upon the concept of paradigm by Kuhn [4]. In a Kuhnian sense, the raising of questions to which the existing paradigm fails to provide adequate answers to is closely linked with the ability of existing scientists, which is ironically built from within the previous paradigm. This ability, although mostly attributed to the conceptual infrastructure, is also closely linked with the physical infrastructure of science, i.e., the instruments and tools that a scientist can access. The true outcomes of the Copernican revolution in astronomy were only realized after the construction of Uranienborg, the first custom-built observatory in modern Europe, and later the invention of the telescope. This enabled not only providing better and alternative answers, but also expanded the opportunity space of questions that can be asked by that era's scientists and philosophers.

With this description of the scientific paradigm in mind, let us construct the concept of the scientific divide in QT with two branches. First is the new scientific tools and techniques emanating from QT. We can expect that an emerging set of QT that relies on quantum information science (QIS) can result in similarly revolutionary impacts of those that were previously developed relying solely on quantum mechanics [1]. The role of quantum mechanics in understanding the behavior of physical objects have already influenced the priorities and research focus in key research areas such as materials science, with an increasing emphasis on designing materials with specific quantum properties for technological applications. A further shift in capabilities among certain scientific areas can be caused by the introduction of new scientific tools and techniques. Common techniques, such as scanning-tunnelling microscopy [46] and various spectroscopic methods that rely on quantum principles already lead to a shift in how experiments are designed and interpreted in many fields of physics and materials science. Furthermore, the utilization of techniques coming from QT already played a role in the redefinition of the kg (SI unit for mass) in 2018 [47] and kickstarting the recent surge in gravitational wave astronomy [48] which resulted in the observation of gravitational waves from two black holes merging and the following awarding of the Nobel Prize in Physics 2017 for LIGO.

The tools enabled by QT have applications beyond physics and chemistry as well. An example is the recent progress in developing magnetoencephalography with optically pumped magnetometers (OPM-MEG) [49] which is a promising tool for the next generation of functional neuroimaging [49]. MEG is a functional neuroimaging technique for mapping brain activity by recording magnetic fields produced by electrical currents occurring naturally in the brain, using very sensitive magnetometers. Current MEG devices require considerable physical infrastructure like cryogenics and superconducting devices, they are similar to MRI devices in their physical configuration. In comparison, the novel OPM-MEG devices are at the size of a bike helmet, a helmet specially 3-D printed for the subject's head with dozens of synchronized quantum sensors placed on it. Enabling the utilization of portable and wearable quantum sensors that can perform MEG operation enhances the opportunity space for questions that can be asked by neuroscientists using these tools for data gathering from subjects in motion. There are already reports dating back to 2018 [50] of OPM-MEG devices being used for research on people playing table tennis, drinking coffee, and performing other physical activities, which were previously not possible due to the limitations of the experimental tools. These techniques can quickly be adopted by teams of scientists using MEG devices for their research, as they already possess the capabilities to work with data generated from these devices. However, according to a market analysis report in 2022 [51] there are only around 200 MEG scanners in use worldwide for both medical and research purposes, which limits the scientific expertise that has any hands-on experience in MEG-based neuroscience to research groups in a handful of countries. These experienced teams will most likely be the first customers of companies producing OPM-MEG devices, have the first access to their capabilities, and be the ones to kickstart a surge of activity in their fields.

The second branch is the adoption of new tools and techniques by different scientific communities resulting in a gap between them in their ability to ask or answer questions arising on the peripheries of normal science. These divisions can be related to funding, representation, educational access, or variations in scientific achievements, among others. Consider the OPM-MEG case. Eventually, the devices will be adopted by others and overall, empower their relevant scientific communities in more countries than currently using MEG techniques today, but this can take varying amounts of time depending on the trajectory of the technology. Therefore, it initially contributes heavily to the creation of a divide in scientific capabilities within certain research communities with respect to their ability to conduct research using these newly emerging quantum tools.

It can be argued that the historical emergence of quantum mechanics made a variety of interdisciplinary collaborations - between scientific fields that were traditionally separate- possible. The roots of the periodic table go back to the mid-19th century, however unlocking the true potential of the periodic table had to wait until the formulation of the Pauli exclusion principle in 1925. Similarly, Schrödinger's ideas on applying quantum principles to understanding the biological concept of life contributed to the discovery of DNA [52]. As has been seen again and again, progress in scientific research is more fruitful in the case of collaboration.

In the case of QT, quantum physics and computer science coming together enabled research into quantum computing, making physicists, engineers, and computer scientists work more closely to build actual quantum computers. Today, the field of QIS, which includes quantum computing, cryptography, and communication, is becoming a distinct field with its own focus and experts [53] and its own discipline-culture [54, 55]. QIS focuses on using quantum properties for processing information and can be considered as the scientific basis for a considerable portion of the QT associated with the second quantum revolution. However, certain approaches and formulations beyond the immediate applications of QIS in physics and computer sciences are not reaching or being accepted by their intended scientific communities. For instance, it is argued in the literature [56] that quantum biology is not being regarded as a "legitimate" scientific field, although it carries significant potential in many areas of biology and life sciences. The reasons for this are identified as the absence of a cohesive quantum biology community, lack of experimental verification, and the continued presence of scientific silos at institutions [56, p.2]. Similar arguments can be applied to the lack of adoption of quantum chemistry techniques using quantum computers for full model simulations by chemists who are sticking with widely available density-functional theory (DFT) modelling, even though quantum computers hold great potential for the field [57]. Another example in the literature is about the potential impacts of quantum computers on research in humanities from data preparation, techniques and methods like clustering, and classification to various approaches for data analysis [40], [41]. Likewise, some authors argue [58] how basic principles of quantum mechanics can be applied to decision-making paradoxes in psychology and used for modelling information in finance and economics. Even between the fields of QT, there are versatile opportunities for further collaboration and development that are not being fully explored, such as examining quantum metrology through the lens of QIS [59].

There are, of course, some who take the risk of exploring these potential new opportunities. For example, according to The Guy Foundation [60], there are almost 50 quantum biology research centers distributed to more than 10 countries, and there are particular examples of strong commitment such as Cleveland Clinic in Ohio signing a 10-year partnership with IBM focusing on advancing the pace of biomedical research through the use of quantum computing [61]. Some of these research efforts will inevitably lead to dead ends. However, it is only expected that successful endeavors will end up providing their research groups with first mover advantage and the opportunity of pioneering previously unexplored topics.

As the scientific divide in the QT context is a nuanced conceptualization, we want to provide two analogical examples related to scientific fields from the era of not the second quantum but the digital revolution, particle physics and library sciences. The ties between particle physics and digital technologies go back quite a while, as it was a field requiring high computational capabilities. For example, some of the most important protocols that enable the World Wide Web (WWW) were first developed by Tim Berners-Lee in 1989 while he was a fellow of CERN, as CERN was the largest internet node in Europe back then. This early adoption of digital technologies, when combined with the resources provided by CERN, enabled particle physics to grow as a truly international scientific endeavor. As an outcome of this, the 2015 article on the first measurement of the newly discovered Higgs boson had 5,154 individual contributors [62], breaking the record for having the largest number of contributors to a single research article. Another example is the field of library sciences and the integration of bibliometrics with the developing digital infrastructure in mid-1990s [63], which also contributed to the development of digital tools, such as Google search. In both cases, tangentially relevant fields which fundamentally have nothing to do with the development of the physical technologies and the infrastructure of the digital revolution, benefitted tremendously and co-produced some of the core functionalities of the digital infrastructure we all use today, such as the internet.

TABLE I: Available/Adopted Scenarios

| Available | Adopted |
|---|---|
| Has access | Uses |
| Has access | Does not use |
| No access | Would use it if had access |
| No access | Would not use it even if had access |

In summary, the two branches of the quantum divide in science we highlight in this section are primarily the availability and adoption of new scientific tools and techniques emanating from QT in general and QIS in particular. The branches can further be diversified, but the main formulation of a gap between those who have potential access to the tools and knowledge and those who do not remain. The second branch adds a further dimension, a gap between those who can realize this potential via early adoption, hence having an impact on the trajectory of that technology, and those who will not. A summary of these four scenarios can be found in Table I. Therefore, the real sizeable gap will expectedly be between two groups: (i) those with potential access to the tools and knowledge, who will adopt these successfully to answer questions on the periphery of normal science, and (ii) those without the potential access to the tools and knowledge who would not realize this potential even if they had the opportunity, hence missing to identify the point in time where these resources will finally become potentially available for them.

*A. II. Quantum Divide in Technologies through Path-dependency*

In all emerging technologies, governments, firms, scientists, and societies play a role in the development of that technology in the quadruple helix framework of innovation ecosystems [8]. From this point of view, it is natural to expect that the interactions between government, industry, academia, and civil society stakeholders impacts the trajectory of an emerging technology. As introduced in Section I, the Kuhnian conceptualization of scientific revolutions and paradigms are reformulated for the field of economic development in the name of the neo-Schumpeterian effort to understand innovation and processes (and difficulties) of technological catch-up in the 1980s by authors such as Dosi [5] and Perez & Soete [6]. This formulation relies heavily on the distinction between invention and innovation, and the fact that "…*the space of the technologically possible is much greater than that of the economically profitable and socially acceptable*" [7, p.186]. Thus, which technology takes the lead in certain areas and for what reasons heavily relies on factors outside of the bounds of that technology's technical capabilities. Societal and economic stakeholders from government, industry, academia, and civil society all play a role in setting the conditions for adoption and trajectory of innovation.

In the context of QT, we make use of another notion from evolutionary economics, path dependence, into this discussion to formulate our theoretical ground. The concept of technological path dependence suggests that the managerial choices made early in the development of a technology can have a lasting impact on the direction and trajectory of research and development in that field [64, 16]. Antonelli defines path dependence as "...*the set of dynamic processes where small events have long-lasting consequences that economic action at each moment can modify yet only to a limited extent*" [64] (p.1). The concept can be operationalized in a way to explain both internal impact and external impact (see Figure 1).

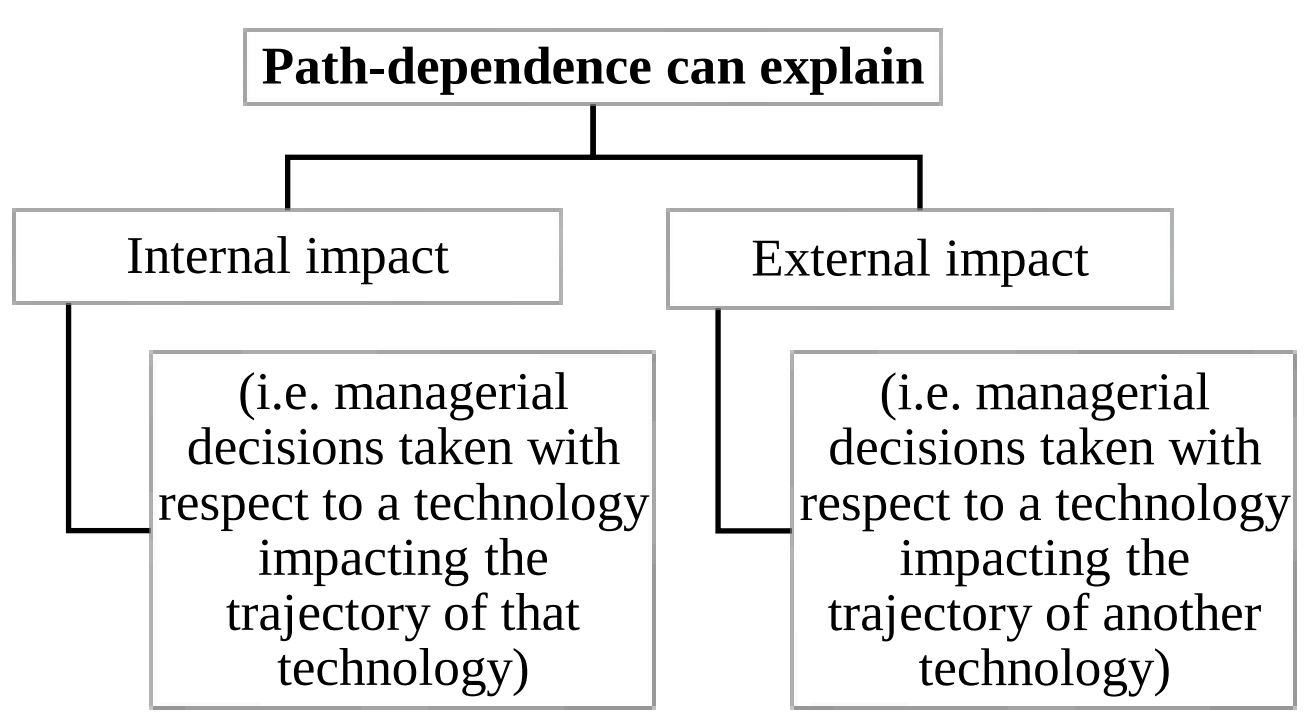

**Fig. 1.** Impacts of path-dependence.

Therefore, we can state that any adoption of a developing or ready-to-use technology by governments, firms, universities or societal stakeholders result in an environment that prioritizes some potential paths while blocking or deprioritizing others and making some other technology's development or adoption easier or harder in the future.

The selection of specific QT by different stakeholders can lead to a diverse global technological landscape, with different countries specializing in various areas [65, 66]. Being able to come up with the dominant design can lead to increased economic competitiveness, technological leadership in QT, unique security capabilities for a while and exerting international influence. Some nations/firms may excel in rapidly developing a first working version of fault-tolerant large scale quantum computers, achieving the first-mover advantage. This can cause the infrastructure to “lock-in” to that certain modality, regardless of the inefficiency of the technology compared to the other possibly more efficient, yet not realized designs. This hypothetical, but well-known process assures the cruciality of the timing [67]. A partially dominant design that leads to at least some favorable returns on investment (ROI) might attract more investment, leading to further improvements, leading to higher ROI attracting more investment and so on. This is referred to as the virtuous cycle [68]. The first to achieve a functioning virtuous cycle can ride the wave of early-stage high growth period of any radical technological innovation. This results in a high switching cost for the later adopters due to most of their previous investment becoming sunken cost depending on when they decide to switch from their modalities to the modality of the dominant design, eventually causing a lock-in where it practically becomes impossible for any other modality to compete with the dominant one [67].

The concept of virtuous cycle for QT is proposed in the literature [68] with respect to how the semiconductor industry aligned itself related to transistors. In this cycle, the adoption of new technology in products leads the industry to make more money, which is then reinvested in the development of newer technology. This is identified [68] as the main driver of Moore’s law, the push for making the number of transistors in an integrated circuit double every two years. It is argued that this virtuous cycle of development in transistors also effectively ruled out alternative technologies for computation via locking-in the information technologies industry to work with silicon transistors for the last 60 years [68] (p.18). Before the invention of the transistor, if you asked a researcher to envision the future of computation, the most probable reply that you get would be miniature vacuum tubes. A similar shift in the dominant modality of physical realization of quantum computing is also possible. Currently, companies like IBM, Google, Rigetti, IQM, and others are focusing on developing superconducting circuits-based qubit systems. However, there are others such as IonQ, Quantinuum, and Universal Quantum working on trapped ion quantum computers, and those such as Xanadu and PsiQuantum working on photonics-based quantum computation. It is entirely possible to look back at the current superconducting circuits-based devices available on the cloud today and consider them as archaic modes of doing quantum computation, as we do for vacuum tubes of the 40s.

In order to evaluate the developmental trajectory [5] of QT, the national initiatives related to QT in various countries and their collaborative efforts on an international scale carry significance. This is because that they don’t only have a direct impact on the quantum divide between countries, but also the policies and funding decisions of governments significantly impact the direction and rate of development in QT. Countries allocate resources and establish policies that encourage research, development, and commercialization in specific QT domains or even on specific modalities for quantum computers, networks, and sensors. Countries generally choose the technologies to fund according to their existing capabilities within their national infrastructures. Moreover, international collaborations as well, are driven by both innovation and risk-averse strategies [69]. By intending to work with countries that have similar approaches, nations aim to minimize the chances of facing unexpected challenges along the development path they are on. Hence, it is only expected that the way countries decide on their pioneer QT, choose which technologies and modalities to support, and collaborate internationally are interconnected.

On a similar line, countries that align on not just dominant design but also values expectedly gravitate towards each other. This is put forward in the literature [70] via the feature of "like-mindedness" in the partnership of the AUKUS security agreement between Australia, the United Kingdom, and the United States, suggesting that having a shared mindset is an advantage of becoming pioneers in the quantum field. This collaborative strategy focuses on strengthening each other's existing systems, instead of exploring completely different paths. Furthermore, the National Quantum Coordination Office of the United States [71] presents international partnerships formed based on the concepts of "shared values/commitments", "common dedication/priorities", and "history of scientific engagement" in such scientific and technological advancements between countries. Under the requirements of international collaborations, whichever technology can be supported the best in terms of material and human capital is likely to be a winner in this pragmatic decision-making process. In other words, shared perspectives, such as a common understanding of determination and values within a governance domain, result in a strategic partnership to work on specific technologies or designs that will support the aimed value and make nations stronger together. Consequently, it is plausible to assert that the values of the nations, which are strong and/or compatible enough to form an alliance, evolutionarily affect the path-dependent process of technology development and adoption.

Combining these, we can expect a quantum divide in technologies to emerge at the point of lock-in [67], not only between countries, but the techno-economic paradigm options beginning to form around different dominant modalities [7], especially for quantum computers and networks. Innovation ecosystem actors of all types in the quadruple helix model [8], academia, industry, government, and civil society, may find themselves on different ends of the spectrum in terms of their involvement with the final dominant modality and system designs. There is a multitude of operational options and room for managerial discretion in the future technological state of QT. Different modalities of implementation for QT can be thought of as technologies where "the adoption of both increases profits by less than the sum of the profit gains from the adoption of both alone" in the path-dependency frame [16, p.7]. This indicates that there will be a market pull towards a tipping point after which lock-in is settled and in a relatively short period of time any previous investment into competing modalities becomes mostly sunken cost. If this adoption is also in alignment with the formation of a new and stronger virtuous cycle, then any new investment into competing modalities also becomes mostly in vain. Leading to an outcome where actors within the helix that invested their time, effort, resources, and attention into anything other than the locked-in modality find themselves in a challenging position. Their situation would resemble having huge investments in vacuum tube manufacturing capabilities in the mid-50s or a movie-renting establishment serving CDs in the age of high-bandwidth streaming services on mobile devices.

*A.III. Quantum Divide between Countries*

There are several aspects that we can consider as the factors causing the emergence of a quantum divide between countries. However, in the context of this article and what is covered in earlier sections, we will limit our analysis to two possible causes for this divide which are summarized in Figure 2.

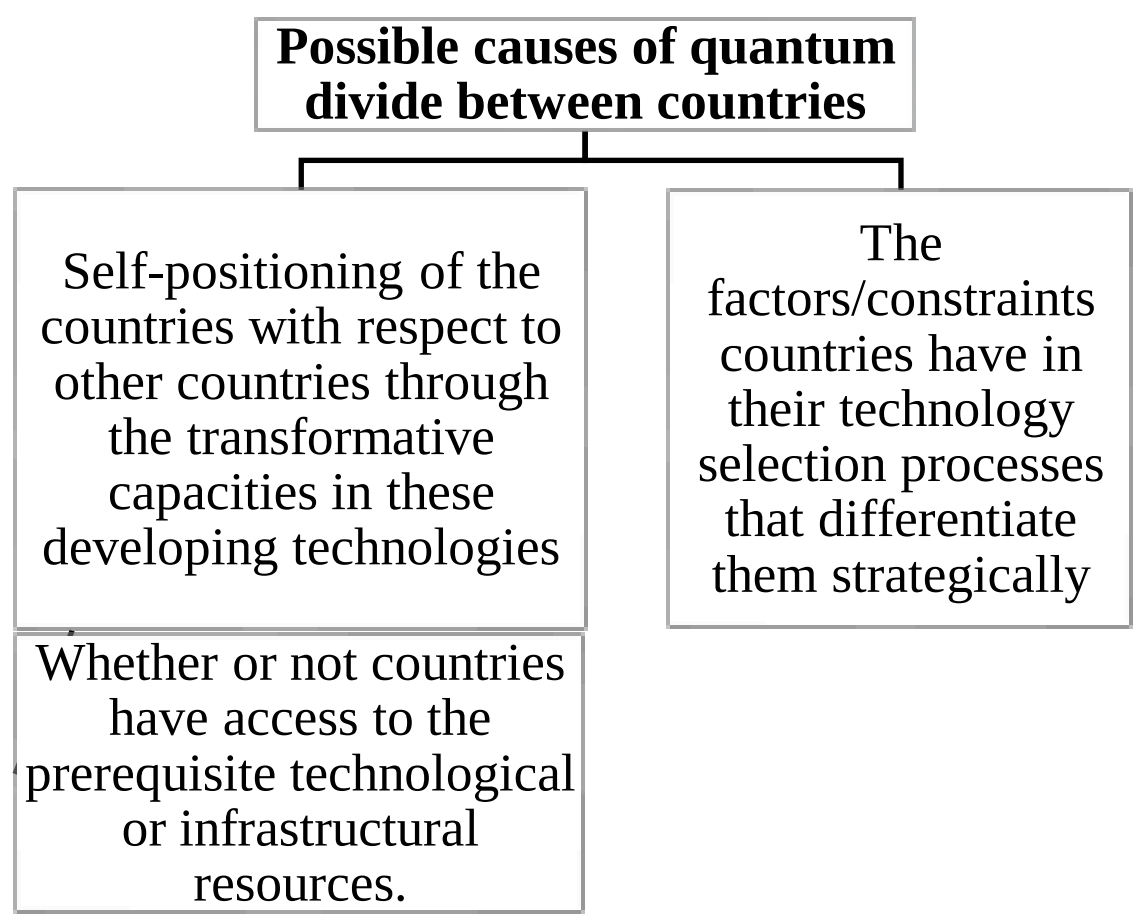

**Fig. 2.** Causes of quantum divide between countries.

As discussed in the literature [72], numerous investigations in the realm of international strategy indicate that each nation establishes its unique knowledge networks, institutional settings, and economic structures, all of which can consistently impact innovation endeavors within domestic markets, no two countries are the same. In determining a nation's technological trajectory, these settings, networks, and structures serve as a starting point and directly affect the country's technological capabilities and options. Nations have various criteria to consider when adopting or developing a technology. These criteria are logically based on both the features of the technology in question, like trade costs and productivity levels, and/or the country's ideology, strategy, and capabilities. Although we suggested that whether there is access to technology or resources is not the only determinant of the divide, it is an undeniable aspect of it. It is included in the capability aspect and carries a historical burden for being a cumulative incident, related intrinsically to the path-dependency [16] introduced earlier. To give an example from the digital divide, a crucial aspect of this divide is the availability of adequate international bandwidth, which is essential for smooth and unrestricted access to the global internet [17]. It is also suggested that lower-income countries often struggle with collecting and effectively utilizing data due to insufficient infrastructure and expertise. This results in a situation where previous decisions, such as lack of investment into infrastructure, a botched attempt at digital sovereignty via overregulation, or educational priorities resulting in an inadequate pool of expertise limiting the options of a country for digital transformation to gap the divide.

As discussed above, the development of new infrastructure is inherently linked to and built upon the foundations of the previous infrastructure. An example from the QT field is that the quantum internet [12] requires a well-functioning 'classical' internet to operate. It will be built upon the existing infrastructure and not replace it. The challenge for QT in this case arises when a seemingly innocent motivation for "self-sufficiency" is pursued without understanding how technology and infrastructure developments are interconnected. Self-sufficiency in strategic areas is usually referred to as a desirable target for many countries [73], and a general definition of self-sufficiency is given [74] as "*a situation in which a country or region's domestic production equals its domestic effective demand*" (p.13). This perspective is adopted by a growing number of nations to emerging technologies to varying degrees because these technologies have a strategic role. It is framed [73] as "technological self-reliance" and is being put forward as an explicit goal by certain countries. For example, China states that it aims for technological self-reliance in its most recent five-year plan [75].

In the context of QT, this aspect is briefly covered in the literature as protectionist impulses [26], especially with a focus on UK's National Security & Investment Act, the multi-state Wassenaar arrangement, and the US's ITAR regulations, which are utilized to support outcomes such as the UK-US Joint Statement on co-operation in QIST (Quantum Information Science and Technology). These can be extended to cover the AUKUS security agreement between Australia, the UK, and the US, and the Commission Recommendation on critical technology areas for the EU's economic security for further risk assessment with Member States [76], which includes QT as one of the recommendations that is already accepted by the EC. Some authors even argue that pursuing these protectionist impulses too strongly might lead to moral isolationism for Europe [77], and is being fueled by an increasing adoption of techno-nationalism [78], an argument on linking a nation's technological capabilities and self-sufficiency to its state security, economic prosperity, and social stability.

These are examples primarily emanating from defense and security concerns. However, a more general shift in the language in recent years is hard to miss in the QT domain. The Quantum Manifesto [79], published in 2016, calls for encouraging international collaboration through new international funding mechanisms in science and promotes international collaboration, exchange, and networking of people and information between different centers in innovation. In contrast, the sub-section dedicated to international collaboration in the preliminary Strategic Research and Industry Agenda (SRIA) document of the Quantum Flagship published in 2022 [80] and laying out recommendations for the roadmap to 2030 is titled "International Collaboration / Export Control Regulation" and international (extra-European) partnerships are recommended in the entire SRIA only within the framework of standardization efforts. Even within the EU, the discussions on going wide or going tall, meaning either distributing the investment and infrastructure to all the member states versus keeping them focused in certain areas/countries for increased efficiency and economies of scale, is an active topic of discussion. Central member states that can support major national initiatives on QT (such as Germany, the Netherlands, and France) are arguing for European support towards their home efforts meanwhile member states with lower levels of infrastructure are calling for support to promote their efforts in catching up.

Naturally, the strategy discussed earlier, which focuses on enhancing current strengths through the selective development of technology, also relates to the issue of the quantum divide among countries. These strengths display themselves in scientific, political, and economic arenas may contribute to a gap among nations based on their advancements in QT. Some in the literature [72] suggests the possibility of an argument that scientists and engineers from technologically advanced countries are more likely to concentrate on studying established technologies rather than exploring completely new directions. This may be due to a variety of factors, including the need to maintain existing infrastructure, the presence of established industry players with significant investments in existing technologies, and the pressure to deliver results. Consequently, in technologically advanced countries, the companies benefit from the positive network effects of well-established technologies, like incubation facilities or infrastructure such as clean rooms, which makes them choose the technology to develop that is supported by the already existing structure, networks, and settings. Essentially, some countries might end up getting better at certain QT, creating a gap between those who are already strong in these areas with their well-established capabilities and those who are comparatively less equipped. Surely, differing perspectives exist on this matter. For example, some argue [81, para. 4] that the development of QT is still in its early stages, "*where it is possible for small players to make a significant impact*". According to this point of view, contrary to the countries with well-established capabilities, "small players" need to identify a specific area of focus that includes originality and high risk within an emerging technology that has the potential to create a worldwide influence. Therefore, it is suggested that [81] small countries not become consumers but creators by taking the risks that bigger countries did not take.

Another topic potentially relevant for QT is the workforce size not meeting the need for labor, which is presented as a bottleneck to quantum development in the literature [82]. In the media, it is portrayed as a national security vulnerability [83] and a battle to train quantum coders [84], and many initiatives around the globe are being implemented to remedy it, with dozens of specialized QIST and QT graduate programs being formed [82, 85]. This is not very surprising considering the number of QT start-ups growing from being less than 20 in 2012 to over 400 in 2022 [86]. This, combined with the point on global mobility above, is a concern for certain countries. To be clear, being an expat or migrant researcher still has many difficulties and hurdles even in the field of QT where fast-tracking visas are being discussed, some of them beautifully described in the paper "Quantum researcher mobility: the wonderful wizard of Oz who paid for Dorothy's visa fees" [87]. However, topics such as depletion of academic reserves and senior brain drain are openly discussed in public forums and sometimes are even causing friction on panels with representatives from the US and the EU having differing opinions on it. Different regions and countries have different priorities in terms of technologies they focus on, modalities they aim to scale up, and levels of risk they wish to accommodate for potential future gain, but they all need a certain amount and type of labor for their efforts. Currently, this labor consists primarily of highly educated technical workers, mostly with PhDs in Physics or certain specialized Engineering degrees, and there are only a limited number of those.

The different priorities highlighted above can come in several shapes and forms and are, of course, subject to change with time. For example, the most up-to-date and detailed report of its time published in 2019 by the National Academies Press on quantum computing [68] focuses on two dominant modalities for quantum computers in Chapter 5, trapped ion and superconducting qubits. It considers photonic quantum computers, neutral atom-based devices, and semiconductor qubits under "Other Technologies." In comparison, as of December 2023, the most well-funded quantum computing start-up (not associated with any major existing tech giant) is PsiQuantum, which is working on photonic quantum computers, while the record on the highest number of logical qubits demonstrated is held by a neutral atom company called QuEra [88], both are from the US where the report was written. This type of uncertainty and rapid changes bring forth additional difficulties in planning forward. The QT landscape has a diversity of modalities being explored simultaneously. For example, in the quantum computing ecosystem, the Austrian start-up Alpine Quantum Technologies (AQT) focuses on trapped-ion quantum devices, the Finnish start-up IQM Quantum Computers specializes in superconducting, the French start-up PASQAL on neutral atoms, and another French start-up Quandela on photonic quantum computers. All these require distinct supply chains and non-transferrable skillsets up to a point, further exacerbating the problem with predicting where workforce bottlenecks may arise, creating additional layers of complexities on the skills shortage front.

We mentioned two causes that may give rise to a quantum divide between countries. The first is how countries position themselves, which we argue is highly impacted by the increasing desire and pursuit of national or regional self-sufficiency and the type of infrastructure they already have access to. Secondly, the factors/constraints they have in their technology selection processes that differentiate them strategically, which are their size, development level, existing path-dependencies, and particularly, the available workforce. A combination of these causes can lead to a considerable divide even within the class of developed economies, where successes in certain regions and countries attract more of the limited workforce there, depleting the academic reserves of other regions, hence further slowing down development in those.

*A. IV. Quantum Divide within Societies*

The final divide we want to explore, which is one of the most highlighted in the literature, is the potential quantum divide to occur within societies. In this section, we primarily want to focus on why this divide is the most expected one and how, in practice, QT can cause the perpetuation of the digital divide into the quantum era.

Technological changes inevitably bring social transformations. They are intricately intertwined, influencing, transforming, and constructing each other in a continuous way. The integration of new technological applications and systems necessarily alters the dynamics of how individuals communicate, work, and live. In turn, societal needs and preferences drive the evolution of technological advancements, creating a cycle in which they feed each other. Following the Schumpeterian claim of innovation causing "creative destruction" [89], one can expect QT to have a profound impact on society.

The meaning attached to technology varies, shaped by common understanding and its usage, creating a diverse foundation for social capital to thrive. Castells' Network Society theory, especially in the realm of digital information and communication technologies [21], emphasizes the transformative role of information networks, particularly the Internet, in contemporary society. It can be inferred from this perspective that the networks that have arisen as a result of significant historical developments - whether from innovations, inventions, or political decisions - have a profound impact. For instance, the emergence of the internet and smartphones has profoundly changed how we communicate and conduct business, altering our daily routines and social interactions. As an outcome of this, we can argue that there emerged an internet culture that is not defined by national borders, but by the commonalities among people regardless of their nations. This cultural shift, driven by technological developments, highlights the potential for QT to contribute to a new cultural order, similar to the transformative impact seen with the internet.

As discussed previously, particular QT, such as quantum computers and the quantum internet [12] will not replace their predecessors, usually referred to as classical computers and the classical internet within the QT community. On the contrary, to operate functionally, quantum computers require considerable computational resources from their classical counterparts. Similarly, for quantum networks to even be able to perform certain operations (such as quantum teleportation), classical communication channels are required. Hence a quantum internet will necessarily require a classical internet to operate on. These facts already hint that the existing digital divide in societies is most likely to be directly translated into the quantum age, only becoming even deeper as not all the ones on the privileged end of the digital divide will get the quantum 'upgrade' to their infrastructure. This upgrade, first, necessitates a physical upgrade, having access to quantum computers or being able to connect to a quantum network from your region. However, as we know from the digital divide literature, having access is only a part of the game [31]. Those with the potential to develop algorithms and applications that can run on these new tools with novel capabilities will require re-training and obtaining new skills. Those who wish to consume or utilize these algorithms or applications will require a certain amount of quantum literacy [90]. The divide will widen between those who have access to these and exploit them in a timely manner and those who do not even have access to the underlying prerequisite infrastructure in the first place.

In a network society, individuals with limited access to digital resources are at a considerable disadvantage compared to those who do in terms of accumulating social capital. A relatively comedic but revealing example of this is social media. Those with access to high bandwidth connection and quality technical equipment have a much higher chance of becoming 'influencers' compared to those who do not. Access to better digital tools enables them to accumulate social capital much more effectively, literally being able to influence those on the other side of the digital divide.

A good example of how access to knowledge and skills related to QT are mediated within societies is which universities have courses or programs in QIS, hence which populations are served. As mentioned in the previous section, there are dozens of new graduate programs that are being created globally in QT [82]. A recent study analyzing the distribution of QIS coursework across 456 institutions of higher learning as of fall 2022 in the US [91] identifies that they are "inequitably distributed, disproportionately benefiting students at private research-focused institutions" and "are largely failing to reach low-income and rural students." This is not a surprising finding, but one that is indicative of the role that QT will probably play when finally, the products of it are introduced to society (such as quantum computers, networks, and sensors). They will primarily be a source for deepening the existing digital divide and will probably create a new one.

## III. Navigating the Divides

Closing the quantum divides is a challenging task, even unrealistic to imagine, particularly given the existing geopolitical landscape that lacks a focus on equitable development for QT. According to de Wolf [20], "...*governments themselves could also have an interest in monopolizing access to quantum computers*" (p.274), which will also affect access to knowledge. He emphasizes the importance of preventing "*more unequal distribution of power and wealth between America and the rest of the world, and between a few big companies and the rest of society*" (p.275). Therefore, although "closing" the gaps is challenging with governments' strategic interests, there is a bottom-up effort, and an ongoing journey to "bridge" the quantum divides.

The concept of navigating divides, as opposed to completely overcoming them, is central to this study's approach. Rather than attempting to entirely eliminate these divides, which may be unrealistic given the structural and systemic factors at play, navigation focuses on managing their impact and creating pathways to minimize their adverse effects. In some cases, navigation involves reducing the width of a divide, as seen in efforts to improve access to quantum research tools or foster international collaboration. In other instances, it requires acknowledging the existence of a divide and implementing measures to mitigate its negative consequences, such as ensuring that marginalized groups are not excluded from the benefits of technological advancements. These approaches are not mutually exclusive but are instead part of a broader strategy that adapts to the unique challenges presented by each type of divide.

Efforts for Responsible Research and Innovation (RRI) aims for accessible, affordable, equitable, inclusive, and understandable technology development and policymaking by "...*opening up conversation between science and wider society..."* [19, p.2]. There are many proposals in the literature on responsible quantum computing and responsible QT, aiming to find ways to bridge the quantum divides [19, 30, 92, 93, 94]. Furthermore, there are suggestions that democratizing access to quantum computing might help fight back against the digital divide as well [27]. Other suggestions in the literature for closing the gaps are interdisciplinary industry and academia collaborations, the role of quantum education, preparing the workforce, public understanding of QT, addressing ethical and security concerns, and having more reflexivity and responsiveness regarding the narratives and actions adopted by the actors in the QT field [95], [96], [97], [27].

In this article, we argue that it is more practical to focus on the navigation of these divides as the initial step because it allows us to include and be aware of the positions of different stakeholders in the field, some of which are, in fact, working towards creating or widening the divides. It is commendable and necessary to think about and act on how to prevent these divides from emerging and aim to bridge them if they do. However, we must also accept the fact that at least some of these divides will emerge in one form or another even if all the actors in the ecosystem wished them not to emerge, and even this is not the case. It is easy to observe that, in nationalist narratives, protectionist impulses extend beyond merely preventing the other party (depending on the perspective of the nation in question); they emphasize the necessity to 'win' the quantum race [98]. This is most clearly encountered in the language of national security and defense, where a strategic advantage, a novel capability that the 'adversary' does not possess, is something particularly sought after. Therefore, we argue that it is worthwhile and necessary to consider exploring scenarios and potential actions that can help the stakeholders of the quadruple helix to navigate these divides, especially those who have managerial duties and whose decisions will have an impact (both internally and externally, see Figure 1) on the innovation processes, as it is more likely than not for them to become a part of our future discussions on democratization of QT [27].

### *A. Navigating the Quantum Divide in Science*

In subsection II.A.I, we explored two potential causes for such a scientific divide to emerge. First, we considered the potential for tools emerging from QT to possess revolutionary qualities that could alter how researchers in disciplines beyond quantum mechanics and QIS conduct their work. Second, we examined the attitudes of different research communities towards the adoption of these new tools and techniques. This dynamic has already been playing out in the field of AI, where it took off following the point where AI was useful as a tool for research communities beyond just those doing AI research. An example of this can be seen in Figure 3. As of 24th August 2024, there are 3321 indexed publications in Web of Science (WoS) with the topic of "artificial intelligence" and category of "Oncology," if we limit the search to only cover the publications before 2015, the resulting number is 68. This corresponds to an almost 50-fold increase in the number of "oncology" papers in the last decade that has the topic of "artificial intelligence." This does not mean that AI research slowed down; on the contrary, it benefited from this adoption tremendously. There were 13,023 WoS indexed publications with "artificial intelligence" amongst their topics in 2014, as of today there are 118,927, an almost 10-fold increase in ten years.

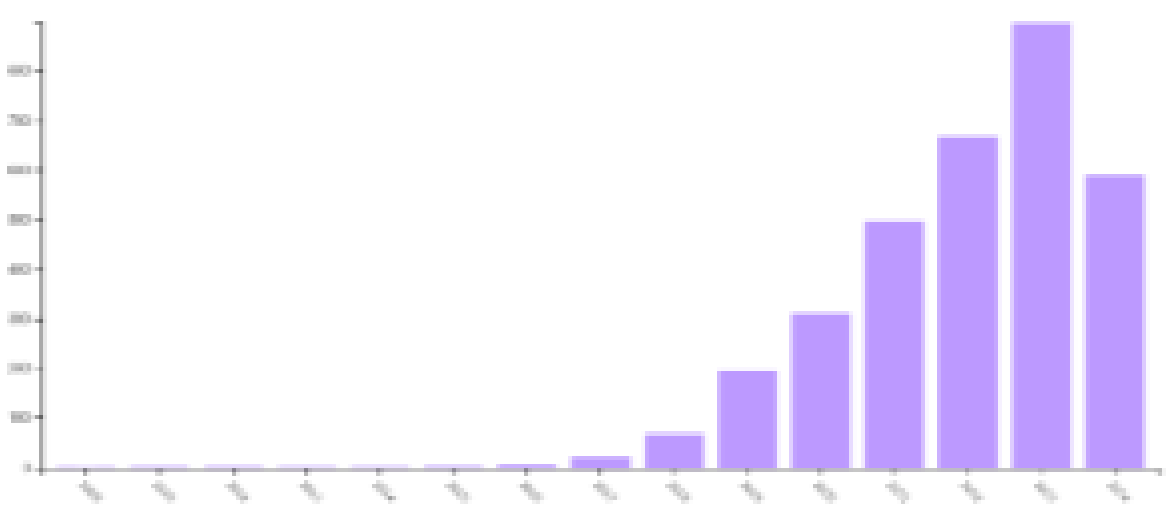

**Fig. 3.** Number of academic publications per year indexed in Web of Science with the topic of "artificial intelligence" and category of "Oncology" as of August 2024

A similar dynamic is expected in the field of QT when quantum computers transition from research tools to practical applications, and quantum networks are established for purposes beyond testbeds and as research infrastructure, serving as either private or public infrastructure, and when quantum sensors are deployed beyond the purpose of testing those sensors. If the promises of QT and the roadmaps of companies like IBM and public initiatives like the National Quantum Initiative Act (NQIA) and the Quantum Flagship are to be successfully realized, we may reach this point by early 2030s in quantum computers, a bit sooner for quantum sensors, and a bit later for quantum networks.

The two ways we propose to navigate this divide in science are relatively complementary, building upon the two causes we argued might result in the emergence of this divide. First, prioritizing the adoption of these novel capabilities by relevant research communities, followed by public and shared research infrastructures with an active effort to introduce them to these communities. The second way is to bring different research communities together, either via creating incentive mechanisms for them to get exposed to each other or by creating joint gathering grounds where boundary objects between disciplines can be formed as their focus of research.

Preliminary efforts that contribute to both these directions are already being carried out by members of all the relevant stakeholder groups. One example from government funding schemes is a recent BMBF grant in Germany titled "Quantum technological and photonic system solutions for challenges of environmental and climate protection, biodiversity, sustainable energy systems and resource conservation" [99], where research not on QT but with QT on other research topics is exclusively targeted. Tailored funding schemes such as these will necessarily require the formation of consortia consisting of parties from different disciplines. Successful project implementation will depend on extensive co-learning within these consortia. International versions of these, such as those via the Horizon program or the bi/tri-lateral research agreements between different countries, similar to a recently announced one between France, Germany, and the Netherlands [100], can contribute greatly to allowing the exchange of expertise between borders as well. An example from industry stakeholders is Deloitte's Quantum Climate Challenge [101] organized since 2021, where Deloitte aims to drive impactful, climate-relevant collaborations between sustainability and climate professionals and quantum computing experts. Similar efforts from different industry stakeholders can also be found, such as the Airbus BMW Quantum Computing Challenge focusing on utilizing quantum computers to forge more efficient, sustainable and safer solutions for the future of transportation. There are also examples consisting of collaboration between multiple stakeholders, such as the formation of the Open Quantum Institute (OQI) [102] and the launching of the XPrize Quantum Applications [103]. OQI is an initiative launched by GESDA (a Swiss NGO), partially supported by the Swiss government, passed along to CERN for operational purposes. XPrize Quantum Applications is a global competition launched by XPrize, Google Quantum AI, and GESDA with a $5 million prize designed to generate quantum computing algorithms that can be put into practice to help solve Sustainable Development Goals (SDG) related problems. Both of these examples focus on using quantum computers for purposes beyond quantum computing research, and both are collaborative efforts between multiple stakeholders of the quadruple helix.

Although the schemes introduced above also help the efforts towards the creation of boundary objects between disciplines, some early examples of boundary objects already exist in the field and can be studied to understand how such objects form. One such object, termed a 'post-quantum signature,' within the intersection of QIS and cryptography is exemplified by the Merkle signature scheme. This scheme, introduced in 2006 [104], was identified as having "a good chance of being quantum computer resistant" at that time. Its significance lies in being the first and most effective one, as it later kickstarted the field of post-quantum cryptography (PQC), resulting in an 8-year long competition by NIST [105] to determine the most suitable PQC algorithms and a declaration by the White House in August 204 that the transition to PQC is estimated to cost U.S. federal agencies $7.1 billion by 2035 [106]. Similarly, another successful boundary object between QIS and computer science, arguably the closest 'outside' discipline to QIS following physics, is called 'a quantum-inspired classical algorithm.' The first example of this appears in the literature in 1996 in the Proceedings of the IEEE Conference on Evolutionary Computation. There are around 1000 articles and proceedings published in WoS indexed sources with the explicit keyword of "quantum inspired" as of August 2024.

As explained in the introduction of this section, neither of these approaches can be expected to fully prevent emergence of a quantum divide in science. However, adoption of these directions can both soften the potential disadvantages of a deep divide, make building bridges over it easier, and overall distribute the benefits of having technological access to the quantum world to disciplines and the pursuit of knowledge in fields beyond quantum mechanics and physics in general. As introduced above, all the different stakeholders of the quadruple helix have options to take action on their own or collaborate with other stakeholders within their innovation ecosystems.

### *B. Navigating the Quantum Divide in Technologies Through Path-dependency*

In the relevant subsection II.A.II, we introduced a potential quantum divide in technologies that may emerge at the point of lock-in [67]. This divide is anticipated not only between countries but also within the techno-economic paradigms [7], with distinct options forming around various dominant modalities, particularly for quantum computers and networks. Such a divide will affect all innovation ecosystem actors who might find themselves on different ends of the spectrum in terms of their involvement with the final dominant modality and system designs.

The way to mitigate the negative impacts of such a divide emerging would be to diversify one's portfolio of investments. However, this is not a task for those who cannot afford diversification and are forced to choose their bets carefully. There are certain actors in the field who can abstract themselves away from this or that modality, focusing on device-independent models of software development approaches or applications that can run on any type of quantum computer. Although this is usually accepted as a good alternative to diversifying for those accepting to not compete in certain aspects of QT (i.e. quantum hardware), it still will not prevent the divide from emerging, as those with dedicated software approaches customized for the final dominant design will have an advantage over those that were hedging their bets.

There is a nuanced point here regarding the future of QT platforms. The potential for this type of divide—where a "winner takes all" scenario leaves others with substantial sunk costs and little to show for it—may lead the entire ecosystem to lock into a technically sub-optimal modality. This risk is particularly high in areas such as energy efficiency, where most players are currently not optimizing their approaches with respect to. The race to deliver any kind of advantage useful for industry purposes motivates the actors not to diversify but concentrate their research and development efforts on already laid tracks. Developing an industry-scale quantum computer is costly and expensive even for the largest commercial actors in the field, such as IBM, Google, and Microsoft, or developing infrastructure-worthy quantum network devices for actors such as IDQ/SK Telecom and Toshiba. The required speed and intensity of investments into these endeavors to deliver outcomes force the hand of these actors to entrench themselves into supply chains and workforce skillsets most suited for the technological path they are most invested in. It is true that "*…the space of the technologically possible is much greater than that of the economically profitable and socially acceptable*" [7, p.186], but it is also true that there are technologically possible innovations which are economically profitable and socially acceptable enough to kickstart a virtuous cycle without being the best option technically. This becomes irrelevant after a certain lock-in point; for example, the trillions of dollars already invested in silicon-based transistors make it practically impossible to switch the entire computing industry to an alternative platform for classical computation, a similar switch that took place in the 1950s from the vacuum tubes to silicon transistors. However, in the field of QT, we are still in the early stages of development. Somewhat paradoxically, the way to prevent this outcome is to slow down the commercialization race, not stifle it, but make it into a marathon rather than a sprint [107] and focus on exploring the opportunity space of possible technologies further. This, of course, will not prevent the divide from emerging between those who are more heavily focused on the final dominant modality and design compared to those heavily invested in another modality with a completely different design. Nevertheless, it will lower the entry barrier to the race and expand the base of those who can afford to diversify. In the end, if we are allowed to abuse the racing analogy, more people can afford to participate in a marathon than a sprint race, and it is not really a big deal to not be the first person to finish a marathon compared to a sprint race.

This is one of the cases we mentioned earlier, where it is in favor of the vested interests of some stakeholders to actually widen the quantum divide instead of narrowing it. Governments and investors that prioritized certain modalities over others and companies developing particular types of quantum hardware would actively benefit from the emergence of this divide and can benefit immensely if it is a wide one. In a plan of action by any of the stakeholders, this conflict of interest should be acknowledged and considered. Bridging this divide via such a diversification introduced above increases the necessary amount of investment into QT to enable exploration of the technologies opportunity space but will result in a more optimal and better dominant design that is also more equitably accessible.

*C. Navigating the Quantum Divide in Between Countries*

In subsection II.A.III, we explored two possible causes for a quantum divide to emerge between countries; (i) how the countries position themselves with respect to other countries and whether countries have technological or infrastructural access to prerequisite resources or not, and (ii) the factors/constraints countries have in their technology selection processes that differentiate them strategically. We primarily explored them through the lenses of self-sufficiency in relation to (de-)globalization [108], existing infrastructure and strengths of countries, the workforce shortage [82], and the corresponding topic of international labor mobility in QT [87]. Out of the four types of divides we describe, this is probably the one that is actually desired and aimed for by a considerable number of actors in the geopolitical arena. Therefore, we believe it is the one that is most likely to emerge strongest and has the most negative impact on those on the wrong side of the divide.

This divide poses unique challenges when considered alongside the other divides within the quadruple helix framework. While the scientific, technological, and societal divides directly align with the stakeholder groups of academia, industry, government, and civil society, the country-based divide operates on a broader scale that often includes actors outside these defined helices. For example, within the United States, governmental agencies involved in quantum technology development fit well into the governance dimension of the quadruple helix. However, the divide between countries such as the US or EU member states and smaller ones demonstrates the limitations of the framework in fully capturing the complexity of global dynamics.

This complexity reflects the nature of the country-based divide as one that extends beyond institutional relationships to encompass disparities between entire nations. While some countries benefit from structured innovation ecosystems embedded in the quadruple helix, others lack the capacity or infrastructure to engage at comparable levels. Despite this, it is important to note that divides also exist within nations, where varying levels of investment, collaboration, and strategic alignment among governmental bodies can mirror the challenges seen between countries.

We will argue for three ways to navigate this divide, but this time they are mostly incompatible with each other. The first one comes from the argument on the small players taking big risks [81]. It is suggested that the small players still have the opportunity to make a significant impact by identifying a specific area of focus that includes originality and high risk, not becoming consumers but creators by taking the risks that bigger countries did not take. This is not a new argument; it actually ties neatly back to the argument of running in a new direction [6] in the (techno-) economic catch-up literature. Through their work in years, developed in the 1980s and 1990s, authors such as Perez, Soete, Dosi, and others identified that for a following country to be able to catch-up to the leading country in a techno-economic paradigm, there are two primary windows of opportunity.

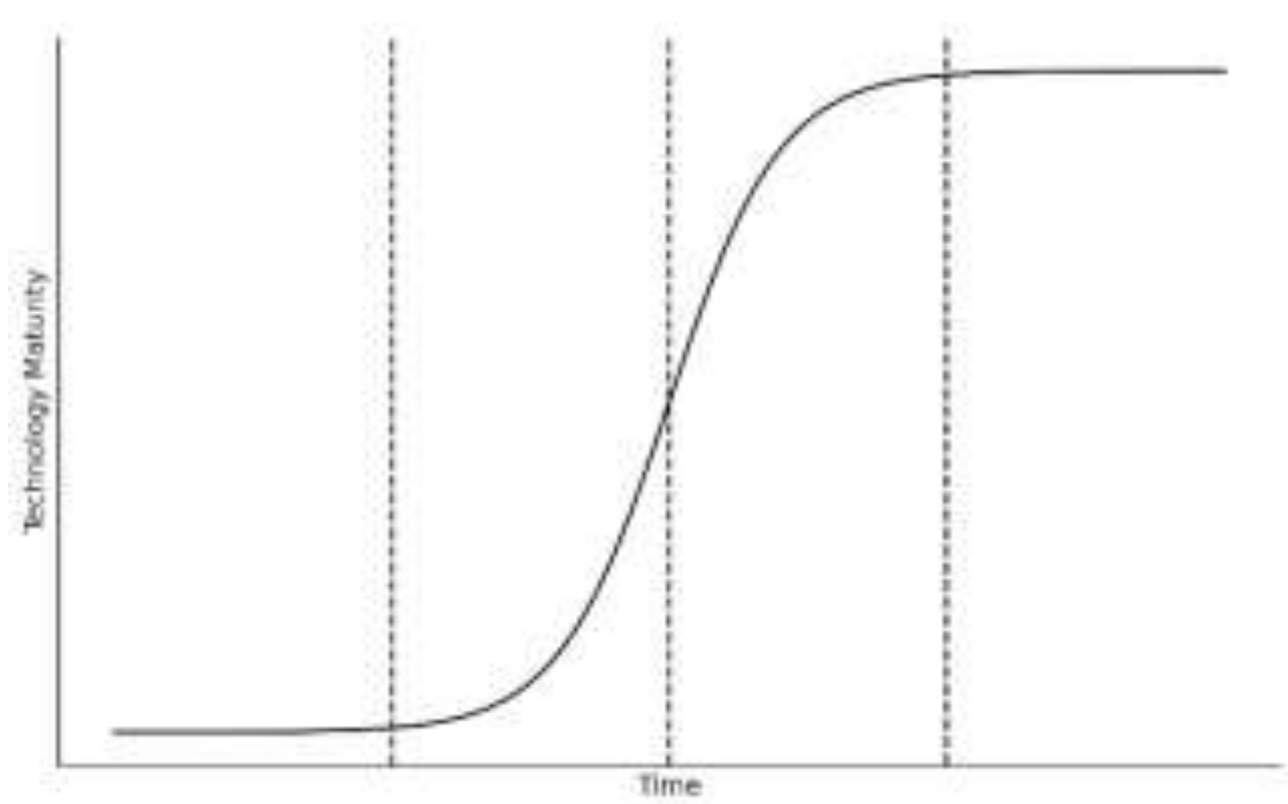

**Fig. 4:** Four phases of technology maturation over time

The easier one to identify is the fourth one, when a techno-economic paradigm is matured, where only incremental improvements are possible (see Figure 4). This creates a period during which new investment into developing a technology further becomes more and more expensive while the return on investment is getting lower and lower, allowing the following countries to compete with the leading country on grounds outside of technical superiority (like labor or land). The second window of opportunity, which is not so obvious, is during the period of a techno-economic revolution, where a following country can choose to run in a new direction and stop following the leading country which is just continuing to make investments into developing a mature techno-economic paradigm to be more productive, getting less and less return on their investment. The argument on small countries taking on risks of running into a new direction which the more established ones are not exploring might actually provide highly beneficial outcomes for them given that they run in the right direction or at least a direction that is good enough to guide through the technological revolution. One successful example of this occasion can be argued as the case of Taiwan and TSMC. The difficulty is that it is not easy to identify such directions, while there are so many different directions that only lead to dead ends with considerable sunken costs.

The second way that can be followed to navigate this divide is to not compete in the race, fail, but fail successfully. Not aiming to catch-up to a techno-economic leader but only aiming to increase the overall prosperity of a country can also be considered as a worthy endeavor. Yet, it would require explicitly accepting the chance of missing out on a valid window of opportunity for technological leapfrogging. It involves acknowledging the role of not becoming creators but mostly consumers of what the technological revolution brings forth. Foregoing the aim of being an early adopter or an early majority, when done intentionally and successfully, this can mitigate the most negative impacts of a quantum divide between countries.

The third way is to team up, especially with regional or established trade partners. A report published by QuEra in July 2024 on survey results of 927 participants from the QT ecosystem titled "The Current and Future State of Quantum Computing" [109]. In which they ask, "*How important is it for your organization where the quantum computer was developed?"* as one of the survey questions. According to their report [109, p4] 27% of the respondents replied that it is important that their quantum computer should be developed in their own country, while 45% of the respondents replied as it is important that their quantum computer should be developed in a country which is a friendly trade partner. This can be used by both the countries on the disadvantaged end of this divide and the countries on the advantaged end of this divide. Ideally, it can be used to match countries from both ends. A good example of this is the EU's QuantERA and Quantum Flagship initiatives, which support quantum research across 31 European countries. When accounted for separately, the countries with the most start-up and commercialization activities in QT are the US, Canada, and the UK. However, when EU-27 is calculated as a single unit, the number of start-ups is on-par with the US for the top spot [75] and the total amount of investment is almost equal to the advertised amount from China. Infrastructure projects such as EuroQCI and EuroQCS cover several countries in the EU, aimed at spreading the QT infrastructure. For example, the six new EuroHPC quantum computers that are going to be integrated into existing supercomputers are located in Czechia, France, Germany, Italy, Poland and Spain, a combination of leading and following countries. Similar initiatives on all shades of this regional and trade partnerships are expected to increase, as there are ongoing preliminary efforts in this direction in regions such as the Association of Southeast Asian Nations (ASEAN) [110] and between allied countries like the US, Japan, and Republic of Korea [111].

We argue that all three ways introduced above are valid ways to deal with the expectation of such a divide emerging. However, it is important to note that these are incompatible approaches. If a country wishes to run in a new direction and try to leap-frog during a technological revolution, all the relevant stakeholders of the quadruple helix [8] of that country should be aligned with that goal. On the other hand, if a country wishes to 'fail successfully' in catching up in a manner that causes the least harm from the emerging divide, its relevant stakeholder communities should act in unison as much as possible. A half-hearted effort to leap-frog will most likely end in considerable opportunity costs, and a half-hearted effort to be a follower might end up causing the country to miss the train and only realizing it when they are very late to the party. Finally, if a collaborative effort is aimed to be put forward, all members should align their goals and priorities with respect to the joint efforts, one member aiming to fail successfully while another aiming to adopt a high risk- high reward approach may end up causing both to not succeed in their respective goals.

One final on this subsection is that identifying the way a country is aiming to navigate this divide is not a straightforward endeavor, as most countries won't have a singular focus on either of the three ways and different governmental agencies with different priorities might push for different ways to navigate. It is important for those with leadership and managerial positions of organizations within different stakeholder groups (such as universities, companies, NGOs, etc.) to keep a close eye on the signals coming from their government agencies on which policies to be pursued and how they are going to be implemented. This is essential for both the government level strategies to succeed and for the organizations in question to benefit from these strategies or at least minimize the harm they would suffer if the policies are designed in a way that are disadvantaging for some stakeholders (like a policy to 'fail successfully' would be detrimental for those already invested in QT startups in a country).

### *D. Navigating the Quantum Divide in Within Societies*

As mentioned in subsection II.A.IV, this is the most highlighted type of divide in the literature, and multiple ways to mitigate the negative impacts and navigate have already been proposed previously. Increasing avenues of access via expanding the user base of these technologies is accepted as a necessary but not a sufficient condition for bridging this type of divide [27]. A renewed urgency to make quantum theory understandable [18], implementing a “strong” RRI approach which entails linking parliamentary or other core policy processes to stakeholder dialogues [30], increasing awareness on both the technologies themselves and their risks [112], creating a dialogue between innovators and societies [113], enabling participation via citizen science initiatives [85], forming and supporting grassroots communities, promoting outreach activities [27], and having value sensitive design practices embedded in the QT development processes [114] are only some of the recommendations made in the literature. However, these discussions are still in their infancy, with only several dozens of articles published on anything related to social and QT as of 2022 [115].

Furthermore, there is a growing number of governance framework proposals in the literature in recent years [27], [116], [117], [118]. The common point of all is the inclusion and empowerment of societal stakeholders, bringing the public up to speed on what QT can deliver, what it cannot deliver, and the wisdom to know the difference. Of course, depending on the theory of democracy one abides to (i.e. participatory, deliberative, or representative [27]), this process can take different forms. For example, in the case of a participatory model, co-creation activities and citizen science approaches might prove to be more essential, compared to a representative approach where civil society organizations and different communities within QT formed as bottom-up initiatives [114] can act on behalf of the societal groups they represent. In terms of a deliberative model, having public dialogue options available where societal stakeholders can connect to these technologies on their own terms [94], [119] would be a good first step to involve them into the deliberation, though requiring many more steps for a truly democratic process.

### *E. Interconnectedness and Distinctness of Divides*

The four divides—scientific, technological, geopolitical, and societal—are not isolated phenomena but interconnected challenges that influence and shape one another. As it is obvious at this point to our readers that there are many overlapping aspects of the different divides we described, there are reinforcing qualities between them where the emergence of one type of a divide increases the probability and potential severity of the other types. There are distinct scenarios as well, where certain types of divides might emerge but not others. For instance, the scientific community could prevent the full emergence of the quantum divide in science by implementing strategies such as intensifying efforts to establish collaborative research infrastructures, actively fostering connections among diverse research communities, and employing other measures. The technology development process might be intentionally (or due to external reasons) slowed down considerably. This can prevent the emergence of a full-blown quantum divide in technologies by giving the relevant actors enough time to phase out of their sunken cost investments and re-orient their resources to join the ecosystem forming around the to-be dominant modality and system design. Countries may decide to give up pursuing a strategic and economic edge through QT, because they might lose faith in the technology or some other possible but less likely alternatives, such as adopting an equitable development model between countries, which should, in theory, prevent the rise of a quantum divide between countries.

While each divide arises from distinct dynamics, their overlaps create opportunities for more comprehensive strategies. A quantum divide in technologies would make investments made towards the ‘wrong’ technology into sunken costs. In practical terms, smaller countries that took risks the larger ones avoided [81] would find themselves not only lagging behind but also heavily committed to a path that most ecosystem stakeholders did not embrace. Such a situation puts a country at a significant disadvantage in the divide between countries. A quantum divide within societies might have an impact on research communities as well. Researchers from fields outside of the immediate vicinity of QIS will have less exposure to QT. Even this limited exposure will be obscured by narratives that present quantum mechanics as incomprehensible [18], further exacerbating the problem of a growing discrepancy between those research communities with access to revolutionary tools and those that do not have a clear understanding of the actual qualities of these tools because their information channels are distorted by narratives mystifying the technology. A quantum divide between countries, when combined with the global mobility of highly educated and skilled labor in the field, could deplete developing economies of their limited number of topic experts to the extent that only a few nations can effectively translate the technical language of QT into a format understandable by the public. This situation can deepen the quantum divide within societies, as the ability to bridge the gap between technical aspects of QT and societal understanding becomes concentrated in a limited number of countries.

By framing the divides as interconnected rather than isolated, the discussion highlights the advantages of an integrated perspective. This approach emphasizes not only the challenges of managing each divide but also the potential for coordinated solutions that navigate multiple divides simultaneously. However, the interconnectedness and distinctness of different types of divides described in this work call for further attention than we can afford in the context of this article. They present opportunities and risks that are not apparent from investigating only a societal, country-level, scientific or technological divide caused by the widespread introduction of QT. The interconnected scenarios also call for navigation strategies considering multiple dimensions at once. This adds further complexity to an issue with already considerable complexity. As the navigation ways proposed in subsections III.A and III.C are clear examples, there are complementary approaches, and there are incompatible approaches regarding the navigation of different types of divides. When considering multiple dimensions, one must also investigate the impact of navigation strategies on other types of divides. If a country adopts a way to mitigate the potential negative impacts of a quantum divide between countries which is incompatible with an approach aimed at navigating the emerging societal divide, then further investigation and deliberation between stakeholders on priorities becomes an urgent issue to handle before taking action.

## IV. Conclusion

In this paper, we aimed to elaborate on the possible divides that could emerge with the development of QT and bring a new perspective on the understanding of how to navigate these divides. We suggested that several types of divides can be formulated. The focus of our study included four divides emanating from the quadruple helix model of innovation, which are the quantum divide in science, the quantum divide in technologies through path-dependency, the quantum divide between countries, and the quantum divide within societies.

Our analysis began with the 'scientific divide,' where we explored how variations in knowledge and resources within the scientific community might lead to disparities in capabilities and opportunities among scientists. This divide underscores the importance of ensuring equitable access to QT knowledge and resources within the scientific arena.

In the realm of technologies, we delved into the concept of path-dependency [16], highlighting how the decision-making process in technology development can lead to certain technologies dominating, thereby potentially excluding others. This creates a competitive landscape where the choice of technology becomes a critical determinant of success or obsolescence.

The divide between countries, perhaps the most geopolitically significant, was examined through the lenses of technological and infrastructural access, self-sufficiency motivations, and strategic positioning in the global landscape. Our findings suggest that the quantum divide is likely to reinforce existing geopolitical and economic disparities, with significant implications for international relations and national strategies.

Within societies, we addressed how QT might perpetuate and deepen the digital divide, transforming it into a quantum divide. This divide is not just about access to technology but also encompasses the ability to understand, utilize, and benefit from QT. The interplay of QT with social structures, norms, and values underscores the societal implications of these emerging technologies.

We discussed that bridging all these divides simultaneously is not likely to happen, and investigation of possible ways to navigate these divides can prove beneficial for all the stakeholders of the quadruple helix, and their respective innovation ecosystems. In this regard, we proposed several ways to navigate these divides. Some of which are complementary and some of which are incompatible to be adopted at the same time, leading to different options and strategies that can be advocated and followed by those in leadership and managerial positions. A summary of these can be found in Table II.

TABLE II: Types of Divides and Associated Navigation Directions Provided in this study

| **Type of Divide** | **Navigation Options (Non-Exhaustive)** |
|---|---|
| *Scientific* | 1. Prioritize and support QT adoption in research fields outside of QT.<br>2. Encourage interdisciplinary collaboration with the QT field. |
| *Technological* | 1. Diversify investment portfolios to mitigate the negative impacts of the eventual technology lock-in.<br>2. Slow down commercialization to explore technology options, lower entry barriers, and give time for ecosystem actors to adjust. |
| *Country-level* | 1. Identify high-risk, original areas for follower countries to become creators, not mere consumers.<br>2. Choose to fail successfully, not competing in the race to catch-up.<br>3. Collaborate regionally or with trade partners to balance the quantum divide. |
| *Societal* | 1. Enhance access and stakeholder engagement in QT.<br>2. Apply varied public participation governance frameworks. |

As we know from the digital divide literature, having access is only a part of the game, and the importance of a technology is found in its integration into the production processes, its role in facilitating information exchange, and its support for engagement [31]. The emergence of QT will impact different stakeholders in a variety of manners; some will benefit from it, some will be just okay, and some will suffer from it. The distribution of these will depend on the paths taken by different stakeholders both within and outside of the QT ecosystem. Neither the benefits nor the harm of any emerging technology is distributed equitably, and the formation of a gap wide enough to be identified as a divide exacerbates this inequitable distribution. The list of navigation ways we proposed in this study is neither exhaustive nor it should be taken as a static list, especially considering that QT has a rapidly evolving landscape. It is essential to note that any way to navigate to be chosen by the stakeholders will require periodic review and updating depending on which of the potential paths in QT are being realized, and how the other stakeholders are choosing and adapting their navigation strategies. It will be a constant game of positioning oneself with respect to others who are themselves with respect to others as well. Exciting times are ahead regarding QT, and it falls on the ecosystem stakeholders to identify and implement ways to minimize the harm of introducing QT to society.

## Acknowledgment

A.A. Gercek and Z.C. Seskir thank Armin Grunwald, Carolyn Ten Holter, and Ceyda Yolgörmez for their valuable time and feedback on the early drafts of the manuscript. A.A.G acknowledges support from the KIT Aspiring Grant. Z.C.S acknowledges support from the DAAD.